\documentclass[runningheads]{llncs}
\usepackage[T1]{fontenc}

\usepackage{graphicx}
\usepackage{tikz}

\usepackage{acronym}
\usepackage{booktabs}
\usepackage{array}
\usepackage{microtype}

\usepackage{xcolor}
\usepackage{hyperref}

\acrodef{SME}{small and medium enterprise}
\acrodef{NCC-IS}{National Coordination Centre for Cybersecurity Iceland}

\newcommand{\quot}[1]{``\kern-0.09em\emph{#1}''} % The \kern is needed because italics slant to the right, moving the top of the first letter away from the upright quotation marks

\begin{document}
\title{Human Factors in Cybersecurity in Icelandic Small and Medium-sized Enterprises}
\titlerunning{Human Factors in Cybersecurity in Icelandic SMEs}
\author{Goda Cicėnaitė\orcidID{0000-0003-1385-4542} \and
Thomas Welsh\orcidID{0000-0003-3134-5469} \and
Helmut Neukirchen\orcidID{0000-0001-8595-3748}}
\authorrunning{G.\ Cicėnaitė et al.}
\institute{University of Iceland, Reykjavik, Iceland\\
\email{\{goc1,tomwelsh,helmut\}@hi.is}}
\maketitle% typeset the header of the contribution
% --- TikZ Overlay for Springer Copyright ---
\begin{tikzpicture}[remember picture, overlay]
  \node[anchor=south, yshift=1.5cm] at (current page.south) {
    \begin{minipage}{\textwidth}
      \centering
      \footnotesize
      \color{gray} % Optional: keeps it visually distinct
      This is the author's version of the work. The final publication will be available at Springer. 
    \end{minipage}
  };
\end{tikzpicture}
% -------------------------------------------
\begin{abstract}
Cybersecurity threats are increasing in all aspects of society due to the integration of digital systems into modern-day life and a volatile geo-political landscape. Technical factors are an ongoing arms race; however, the threat surface from human and social factors is still present, often providing malicious actors the means to bypass complex technical security controls. Understanding human factors in light of technical evolution is essential to ensure security controls remain effective. 
This study presents the results of a survey on cybersecurity challenges within public and private sector organisations, including critical infrastructure providers, in Iceland ($N=130$). From the management perspective, human factors were strongly noted as challenges and barriers to their organisations' security. These challenges include a lack of adequate training or awareness, hiring issues, poor cybersecurity culture, and time and/or financial resource constraints. Based on these findings, recommendations for mitigating threats from human factors are derived. These include: prioritising targeted over generic training to reduce employee fatigue, external government support for financially constrained organisations, and building a strong cybersecurity culture through constructive communication around shared responsibilities. 
\end{abstract}

\section{Introduction}

Cybersecurity is not merely a technical matter; it is also impacted by human factors~\cite{furnell2017security}. Decision-making processes, human interactions as well as organisational culture can affect digital security systems' safety performance~\cite{khadka2025human}. Human factors are well studied for their ability both to enhance and to degrade cybersecurity.

Social engineering attacks, whereby an attacker manipulates psychosocial phenomena to deceive a legitimate user into granting access to a system, are pervasive. In a cybersecurity context, they date back to the 1960s phone `phreakers'~\cite{hatfield2018social} and continue to this day. There is now a diverse array of social engineering techniques~\cite{birthriya2025comprehensive}, further evidencing the effectiveness of attacking the human factor and highlighting the variety of prevention approaches necessary for each attack vector. Such approaches largely consist of training and awareness programmes. Yet, technical countermeasures such as machine learning-based phishing detection and browser-based defences are also present. However, some authors argue that to be successful, these approaches must be informed by the users' psychological factors or organisational context, as cybersecurity is an inherently social phenomenon~\cite{mcalaney2020cybersecurity}. Human-centric cybersecurity frameworks can integrate psychological resilience, socio-technical approaches, adaptive training, and diverse practical strategies to improve organisational compliance, decrease vulnerabilities, and enhance cybersecurity awareness~\cite{khadka2025human}. Even so, for such approaches to be successful, it is first important to understand the context, as society and culture may impact cybersecurity maturity at a nationwide level~\cite{creese2021social}.

This paper focuses on the nation of Iceland and seeks to address the principal research question: What are the main cybersecurity challenges of Icelandic \acp{SME} and public organisations? 

As an island nation, Iceland is a geographically isolated country with a relatively small (around 400,000 inhabitants~\cite{StatisticsIceland}), traditionally homogeneous population with a high level of interpersonal trust and a high level of digital literacy~\cite{ITU2024IDI}. This high level of trust was identified as not conflicting with cybersecurity practices among the general population who accept a national electronic ID as a two-factor digital security solution~\cite{stefansson2024understanding}. However, complicating factors include the potential impact of increased migration on cultural norms and ongoing geopolitical security threats. These factors make understanding how Iceland’s public and private sector organisations are positioned in terms of cybersecurity increasingly important, yet heavily understudied.
To close this research gap, this study investigates the current state of cybersecurity in Iceland, by surveying 130 \acp{SME} via a questionnaire focused on the challenges and barriers they face. This paper presents a subset of these results, focusing upon the human factors.

The remainder of this paper is structured as follows: Related work is reviewed in Section~\ref{sec:Related Work}. The survey methodology is detailed in Section~\ref{sec:Methodology}. The survey results are provided in Section~\ref{sec:results} and subsequently discussed in Section~\ref{sec:discussion}. Finally, Section~\ref{sec:summary} concludes this paper. Data on statistical significance is provided in an appendix.

\section{Related Work}\label{sec:Related Work}
This section positions our work by presenting and discussing selected related studies that focus on cybersecurity in SMEs across different nations.

Colabianchi et al.\ conducted a Delphi study, consisting of 30 participants, to identify managerial actions that can improve cybersecurity~\cite{colabianchi2025transforming}. Their panel consisted of academics with interests in human factors in cybersecurity, in addition to participants from a wide spectrum of industrial roles who have been involved in cybersecurity in some capacity. The participants agreed upon 16 actions which include training / awareness campaigns, correct role management, and appropriate incident response. They also highlighted atypical factors such as ensuring a good workload balance, developing a cybersecurity culture, and encouraging personal responsibility, which indicates the direct impact of the general work culture on cybersecurity in organisations.

Other works explicitly focus on \acp{SME}, which lack the resources of larger organisations to effectively deal with cyber threats. Chidukwani et al.\ conducted a literature survey on studies which focus on cybersecurity challenges in \acp{SME}, highlighting that \acp{SME} are often overlooked. Furthermore, they note that the bias in most studies is towards the United States of America. In addition, the studies themselves are biased towards strategies and policies, yet are biased against implementation, detection, response, and recovery. Consequently, they conducted a survey of \acp{SME}' cybersecurity preparedness across Western Australia~\cite{chidukwani2024cybersecurity}. Despite contacting all 235,060 \acp{SME} in Western Australia, they received only 36 responses, with 29\% of respondents abandoning the survey, likely because of its length. The three main challenges observed were a lack of funds, of cybersecurity knowledge, and of general understanding of cybersecurity standards. 

Khan et al.~\cite{khan2025investigating} take a different perspective by performing semi-structured interviews with 12 organisations that provide cybersecurity support to \acp{SME} in the UK. A key issue they identified was related to communication that was deemed ineffective due to misunderstood technical language and negative messaging associated with the topic of cybersecurity. In addition, the authors note a general lack of resources (financial, human) to provide adequate support. They also note issues with being able to support the wide range of tools and technologies used by the \acp{SME}. In a related study, the authors conducted semi-structured interviews with 12 SMEs and 12 cybersecurity providers to investigate specifically how cybersecurity guidance material (e.g. online resources) can impact their cybersecurity adoption~\cite{khan2025hidden}. The findings highlight that a lack of knowledge, and, as in the previous work, comprehension due to the technical language used, was a common barrier to receiving cybersecurity guidance. 

Kappe et al.~\cite{kappe2023cybersecurity} also conducted semi-structured interviews ($N=154$) with \acp{SME} in Germany with a focus on the relationship between cybersecurity and cybercrime. A key finding related to human factors was that 40\% provided cybersecurity awareness training yearly, 37\% monthly, 7\% weekly, and 17\% only once; indicating wide variation in cybersecurity awareness policy. They also highlight that time and the ability to comprehend cybersecurity information are key barriers to receiving assistance from third parties.

Kabanda et al.\ studied \ac{SME} cybersecurity perspectives in South Africa~\cite{kabanda2018exploring}. They performed semi-structured interviews with employees in varying roles of three \acp{SME}. Key challenges to cybersecurity they identified internal to the organisations were: budget constraints, management issues, and a lack of employee support or commitment to cybersecurity in general. In addition, the authors highlight external challenges, such as the need to adhere to government policy and regulation or professional standards. They note that \acp{SME} feel less concerned with professional standards or training methods than larger organisations which typically feel coerced as part of professional groups. Conversely, these \acp{SME} are more likely to adopt cost-effective open source solutions which have a weaker element of influencing behaviour towards professional standards.

Rawindaran et al.\ study how cybersecurity for \acp{SME} is impacted by support from government~\cite{rawindaran2023perspective}. They conducted a survey of $N=34$ selected \acp{SME} from diverse industries within Wales. Their findings for how the government can support \acp{SME} include the following: providing general guidance and practices, delivering training and offering financial support, performing audits, providing certification, and providing partnerships with cybersecurity firms to tailor offerings towards the \acp{SME}.

This reviewed literature highlights several key findings related to human factors and cybersecurity for \acp{SME}; these include resource constraints related to people, cost and time, as well as a general lack of cybersecurity knowledge preventing information from being understood and consumed. Moreover, organisational and management culture were highlighted as a barrier. Considering that employee awareness was also an issue, yet training was lacking or inconsistently applied across organisations, this might indicate a rift between employees and managers regarding who should take responsibility. The work in this paper contributes to this literature by providing a study of challenges and barriers to cybersecurity for \acp{SME} in Iceland, aiming to understand how these issues manifest in this specific national context.

\section{Methodology}\label{sec:Methodology}

The data used in this study were collected via an online survey based on closed-ended and open-ended questions designed by the Computer Science Department of the University of Iceland. The survey aimed to assess cybersecurity challenges and practices of \acp{SME} and critical infrastructure providers in Iceland as a part of an initiative by the \acf{NCC-IS} to address national cybersecurity challenges. 

The survey comprised 14 questions and 8 subquestions distributed across 6 sections: Organisational background, Challenges, General, Skills and training, Recruitment, Incidents.  
Data collection for the survey was conducted by the Social Science Research Institute of the University of Iceland from  August 13 to September 10,  2025. 641 \acp{SME} were randomly selected, stratified by sector to ensure proportional representation of the Icelandic business landscape. 
To ensure a sufficient representation of critical infrastructure, the sample was subsequently supplemented with 52 critical infrastructure organisations. 

A two-step approach was then used for data collection. The first step included telephone inquiries to the selected organisations, to identify lead staff members responsible for cybersecurity within those organisations. In the second step, the identified lead staff received an email containing the URL of the survey. 

The survey findings demonstrate statistical significance, based on a Chi-Square Goodness-of-Fit test~\cite{agresti2018introduction} (see Appendix~\ref{appendix:chi_square} for details).
This indicates a statistically significant meaningful preference among the respondents, rather than a random distribution of opinions. 

Since the survey included closed-ended and open-ended questions, the RStudio~\cite{RStudio2026} software was used for the statistical analysis of the data and the creation of the figures and tables, the visual representation of data that allowed identifying patterns and comparisons. The data from the open-ended questions of the survey were thematically analysed using Atlas.ti~\cite{ATLASti2026} software. First, the answers were inductively coded by one person using the bottom-up approach; the codes were organised by patterns and themes. Subsequently, the patterns were observed between these established themes. The themes were then checked against the data and codes, a step that was necessary to identify the defining patterns of meaning that are the most important for the study. Thematic analysis provides flexibility and supports the researcher's active role in the process of data analysis through the development of codes and themes  \cite{clarke2017thematic}.

\section{Survey Results}\label{sec:results}

A total of 693 Icelandic organisations were contacted, yielding 130 survey participants (19\% response rate). Among these respondents, 109 completed the survey in its entirety, 4 completed at least half, but not all, of the questions, and 17 completed fewer than half. It should be noted that the initial pool of 693 contacted organisations, and the 130 final respondents include the subset of 52 critical infrastructure organisations, which yielded 20 responses (39\% response rate).

\begin{figure}[!b]
    \centering
    \vspace*{-4.0ex}
   \includegraphics[width=0.9\textwidth]{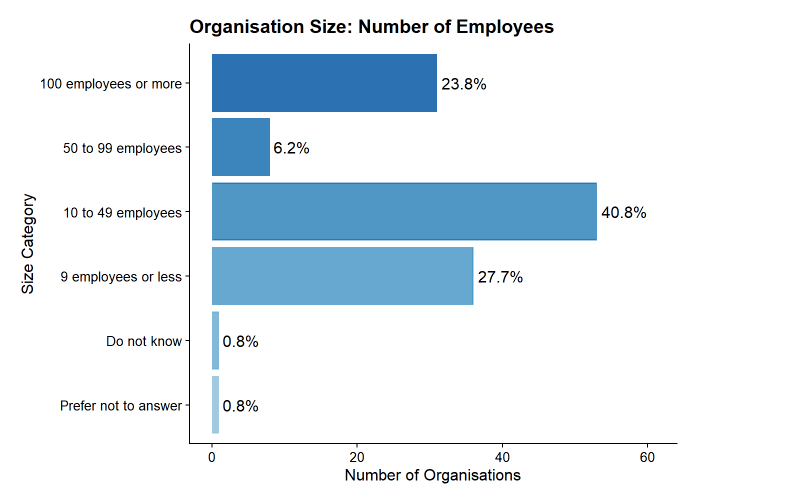}
   \vspace*{-3.0ex}
   \caption{Organisation size (Q1 How many people work in your organisation?)}
   \label{fig:Q1 Organisation size}
\end{figure}

Figure~\ref{fig:Q1 Organisation size} shows the distribution of the organisation size (Q1): 
As the focus was on \acp{SME}, the largest share (41\%) of organisations had 10 to 49 employees, next was an organisation size of 9 employees or less (28\%).
In the responses, the following sectors (Q2) had the largest share: 10\% hotels/accom\-modation/restaurants, 10\% specialist and scientific activities. Two-thirds (66\%) of the survey respondents answered that they are in IT-specific roles (Q3.1), and 83\% of the respondents answered that they are in a senior or leadership position within their organisations~(Q3.2). 
A thematic analysis of the open-ended questions (see Section~\ref{sec:Methodology}) helped to cluster the data into the following themes: cybersecurity management roles, challenges and barriers, perceived protection and incidents, training, and hiring.

\subsection{Cybersecurity management role}\label{sec:Cybersecurity management role}

\begin{figure}[!t]
    \centering
    \vspace*{-3.0ex}
    \includegraphics[width=0.8\textwidth]{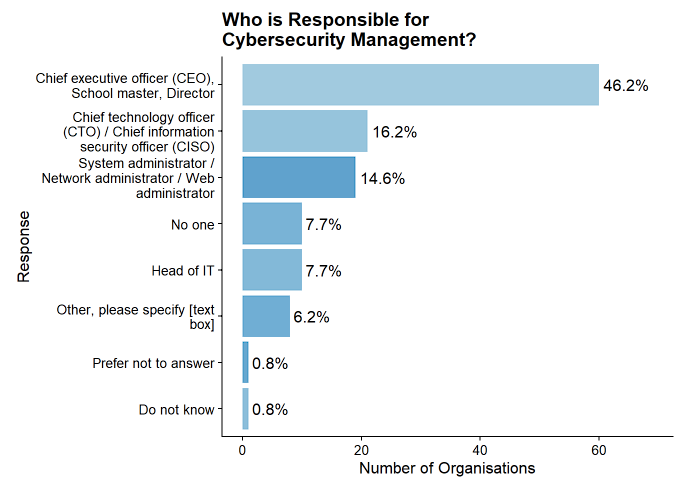}
    \vspace*{-3.0ex}
    \caption{Responsibilities (Q4 Who is responsible for cybersecurity management in your organisation?)}
    \label{fig:Q4 Responsibilities}
\end{figure}

Figure~\ref{fig:Q4 Responsibilities} shows what role is responsible for cybersecurity management (Q4):
the largest share (46\%) of the respondents said that the CEO, school master, or director is responsible. 

\subsection{Cybersecurity challenges and barriers}\label{sec:Cybersecurity challenges and barriers}

\begin{figure}[!t]
    \centering
    \vspace*{-1.0ex}
    \includegraphics[width=0.8\linewidth]{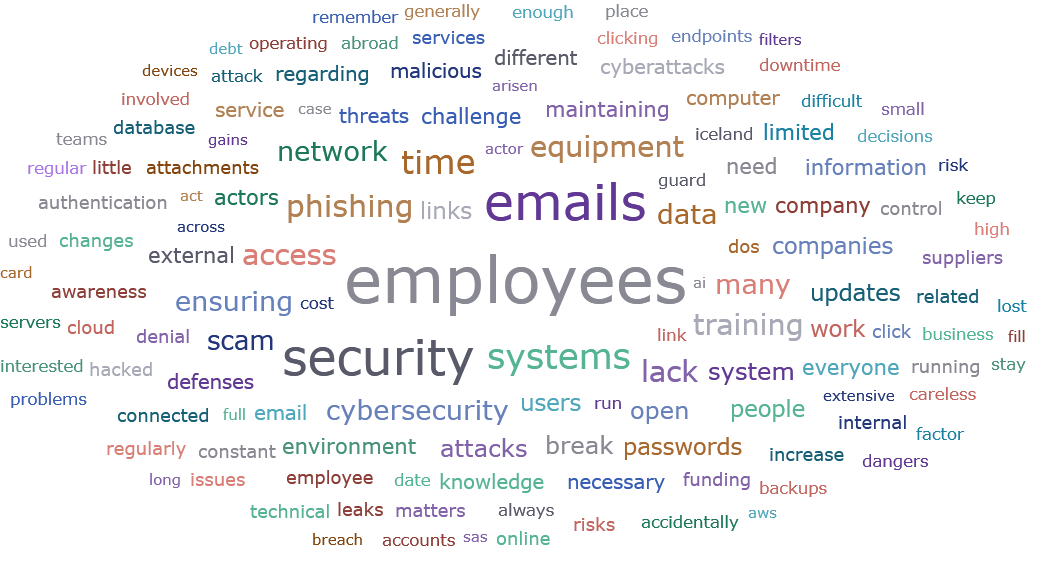}
    \vspace*{-3.0ex}
    \caption{Word cloud for answers to Q5.1 (cybersecurity challenges) and Q5.2 (barriers)}
    \label{fig:wordcloud}
\end{figure}

The main goal of the survey was to identify the main cybersecurity challenges respondents' organisations face. Therefore, Question~Q5.1 asked \quot{What are the main challenges when it comes to cybersecurity within your organisation?}, followed by \quot{Please explain why it is a challenge, list the barriers.} (Q5.2). A word cloud of the given answers is shown in 
Figure~\ref{fig:wordcloud}. Note that \quot{employees} was the most common word, emphasising managerial perspectives on employee responsibility.

Both of these questions were open-ended allowing study participants to write down their answers in a text box. In the responses to Q5.1, the most commonly typed-in challenge was about \quot{phishing} and \quot{emails}, i.e.,\  \quot{careless employees} clicking on scam emails. Other often mentioned challenges include \quot{ineffective} or \quot{unfocused} employee training and the challenge to ensure that employees have digital security knowledge. Further, the written answers to Q5.1 included challenges such as \quot{keeping software up-to-date}, password management and multi-factor authentication considered annoying by staff, \quot{high load of attacks}, \quot{DoS} (Denial of Service attacks), challenges caused by the \quot{need to assure that external service providers have their cybersecurity in place}, and the challenges caused by \quot{credit card or payment frauds}. 

The survey's open-ended answers on barriers to cybersecurity (Q5.2) included \quot{fragmented/too complex IT ecosystems}, i.e.,\ too many different software and hardware systems to manage. 
As the two questions, Q5.1 and Q5.2 are related, some respondents mentioned as barriers what others had mentioned as a challenge: \quot{human errors}, \quot{lack of employees' knowledge}, reluctance to follow efficient cybersecurity practices, unwillingness to spend time on training or updates. 
Further barriers listed by the respondents were economic barriers in the form of \quot{lack of money and/or time}; these can also be classified as both challenges and barriers. These economic barriers also hint at conflicts around responsibilities and organisational culture in general, such as: \quot{Owners and managers generally want everything to be secure, but are not necessarily willing to bear the cost of having the staff required to ensure that this is the case}.

Some of the survey respondents also mentioned the linguistic factors that might affect non-Icelandic employees, pointing out that numerous scam emails received are written in the Icelandic language, and thus some respondents perceive it to be more challenging for their employees with a foreign background to identify scams before clicking on them.

\subsection{Perceived protection, real incidents and protection measures}\label{sec:Perceived protection, real incidents and protection measures}

In question Q6.1 \quot{Do you feel your organisation is adequately protected against the following attacks?}, survey respondents were asked to select on a Likert scale to which extent they feel their organisations are protected against listed attacks. 
Table~\ref{tab:q61 perceived protection} shows that a large majority perceives that they are either to a great extent or to some extent protected against the listed attacks.

\begin{table}[!t]
    \centering
    \footnotesize 
    \renewcommand{\arraystretch}{1.2} 
    \setlength{\tabcolsep}{3pt} 
    \caption{Perceived Protection Against Cybersecurity Attacks (Q6.1)}
    \label{tab:q61 perceived protection}
    % Changed 'p' to 'm' for vertical centering
    \begin{tabular}{>{\raggedright\arraybackslash}m{3.0cm} *{6}{>{\centering\arraybackslash}m{1.2cm}}}
        \toprule
        \textbf{Attack} & 
        \textbf{To a great extent} & 
        \textbf{To some extent} & 
        \textbf{To a small extent} & 
        \textbf{Not at all} & 
        \textbf{Do not know} & 
        \textbf{Prefer not to answer} \\
        \midrule
        Attacks against network equipment & 31.0\% & 38.1\% & 11.5\% & 3.5\% & 15.9\% & 0.0\% \\
        Attacks against web-based services & 23.9\% & 45.1\% & 12.4\% & 5.3\% & 13.3\% & 0.0\% \\
        Authentication-related attacks & 29.1\% & 40.9\% & 10.9\% & 3.6\% & 15.5\% & 0.0\% \\
        Denial-of-service attacks & 27.6\% & 34.5\% & 13.8\% & 6.0\% & 17.2\% & 0.9\% \\
        Hardware/software supply chain attacks & 14.7\% & 43.1\% & 10.1\% & 3.7\% & 26.6\% & 1.8\% \\
        Ransomware attacks & 23.5\% & 47.8\% & 7.8\% & 7.8\% & 12.2\% & 0.9\% \\
        Social engineering attacks (phishing, scams) & 28.7\% & 40.9\% & 20.0\% & 3.5\% & 6.1\% & 0.9\% \\
        \bottomrule
    \end{tabular}
\end{table}

\begin{figure}[!t]
    \centering
    \vspace*{-3.0ex}
    \includegraphics[width=0.8\textwidth]{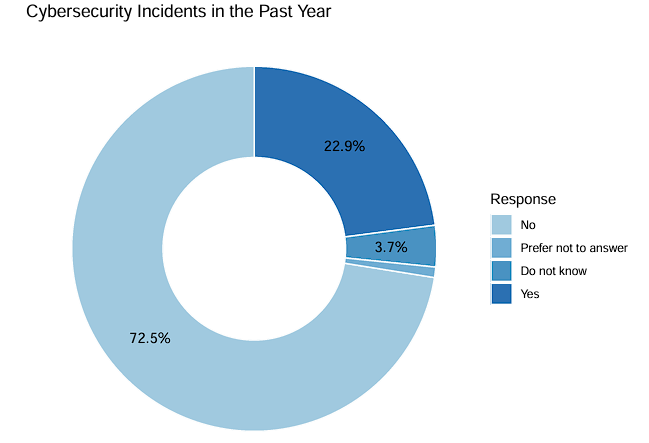}
    \vspace*{-3.0ex}
    \caption{Incidents (Q14.1 Has your organisation experienced any
cybersecurity incidents in the past year?)}
    \label{fig:Q14.1 Incidents}
\end{figure}

However, in another question (Q14.1, see Figure~\ref{fig:Q14.1 Incidents}) almost a quarter, 23\%, responded that their organisation experienced a cybersecurity incident in the past year (i.e.\ 2024/2025). 
These real incidents contradict the perceived feeling of protection.
When asked about the attack vector used for the reported attack (Q14.3), social engineering (e.g.\ phishing) was the most commonly (36\%) selected attack vector by the survey respondents, underlining the relevance of investigating human factors in cybersecurity in Iceland.

Another indicator of overestimating their own protection is the results of question Q8 that asked \quot{How often do you have any kind of security audits? Examples: penetration testing, code review, compliance audit.} 37\% selected a response option
identifying that security audits in their organisations never take place. On the other hand, at almost half (49\%) of
Q8 respondents’ organisations, security audits are performed either at least once a
year or less often than once a year.

\begin{table}[!t]
    \centering
    \footnotesize 
    \renewcommand{\arraystretch}{1.2} 
    \setlength{\tabcolsep}{3pt} 
    \caption{Perceived Protection Against Threats (Q6.2)}
    \label{tab:perceived_protection_threats}
    \begin{tabular}{>{\raggedright\arraybackslash}m{3.0cm} *{6}{>{\centering\arraybackslash}m{1.2cm}}}
        \toprule
        \textbf{Threat} & 
        \textbf{To a great extent} & 
        \textbf{To some extent} & 
        \textbf{To a small extent} & 
        \textbf{Not at all} & 
        \textbf{Do not know} & 
        \textbf{Prefer not to answer} \\
        \midrule
        Insufficient backup of data & 33.3\% & 31.4\% & 14.3\% & 14.3\% & 6.7\% & 0.0\% \\
        Lack of keeping IT systems up to date & 28.7\% & 34.3\% & 13.0\% & 14.8\% & 9.3\% & 0.0\% \\
        Loss of laptops, access cards, door keys & 19.8\% & 40.6\% & 16.0\% & 17.0\% & 6.6\% & 0.0\% \\
        Shadow IT use by employees & 12.4\% & 30.5\% & 24.8\% & 19.0\% & 13.3\% & 0.0\% \\
        Users not adhering to security policies & 15.2\% & 41.0\% & 23.8\% & 11.4\% & 7.6\% & 1.0\% \\
        Weak passwords & 25.5\% & 32.1\% & 22.6\% & 14.2\% & 5.7\% & 0.0\% \\
        \bottomrule
    \end{tabular}
\end{table}

Question Q6.2 \quot{Do you feel your organisation is adequately protected against the following threats?} was about internal threats to cybersecurity. Table~\ref{tab:perceived_protection_threats} shows that a large majority perceives that they are either to a great extent or to some extent protected against the listed threats. However, with respect to `shadow IT' use by employees, the respondents were not as confident that they would be protected as for the other threats.

\subsection{Training employees}\label{sec:Training employees}

\begin{figure}[!t]
    \centering
    \vspace*{-3.0ex}
    \includegraphics[width=0.8\textwidth]{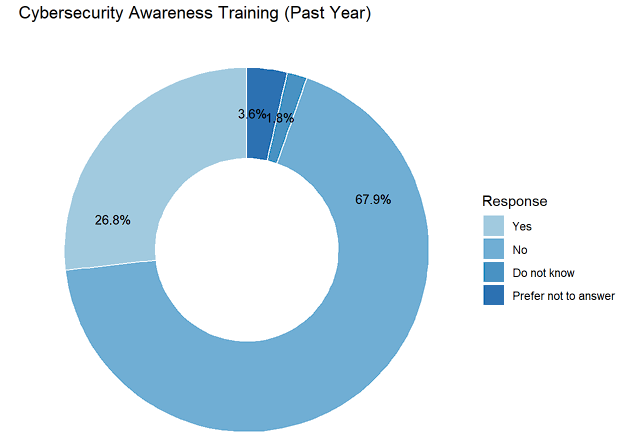}
    \vspace*{-3.0ex}
    \caption{Training (Q9.1 Has your organisation offered any sort of mandatory cybersecurity awareness
training for employees in the past year?)}
    \label{fig:Q9.1 Awareness training}
\end{figure}

Figure~\ref{fig:Q9.1 Awareness training} demonstrates the answers to the question Q9.1 \quot{Has your organisation offered any sort of mandatory cybersecurity awareness training for employees in the past year? (Examples: phishing simulations, educational seminars)}. The data indicates low levels of cybersecurity awareness training in \acp{SME} in Iceland. Only 27\% of the respondents stated that their organisations offered cybersecurity awareness training for employees in the ``past year'' (2024/2025).  More than two-thirds of the respondents, 68\%,  answered that such training was not offered.  

This confirms what was mentioned as one of the main challenges (Q5.1) and barriers (Q5.2) in Section~\ref{sec:Cybersecurity challenges and barriers}, namely insufficient digital security practices and knowledge of employees. Ironically, training of employees is among the most frequently mentioned issues the survey respondents listed as challenges and barriers when answering Q5.1 and Q5.2. Some of the survey respondents also expressed uncertainty whether training provided is \quot{sufficient} and disappointment that \quot{training employees in cybersecurity is time-consuming, costly, and results are uncertain}.

The open-ended survey responses to Q9.2 about the type of awareness training gave insights into the types of courses, tools and platforms used for cybersecurity awareness training. The responses included simulated cyber attacks, trainings with the focus on phishing, courses on secure coding, usage of password managers and general cybersecurity awareness courses for all employees, yet emphasising training provided to new employees by local or foreign providers. Some of the written responses to Q9.2 suggest that \acp{SME} in Iceland use various providers of cybersecurity courses, either local or foreign. Moreover, in the written answers, a clear recurring emphasis is placed on the requirement for all employees to take courses, in some answers, the emphasis is placed more on the training of \emph{new} employees.
The answers to the survey suggest also that training to hypothetical cybersecurity incidents could be improved.

\subsection{Hiring cybersecurity experts}\label{sec:Hiring cybersecurity experts}

\begin{figure}[!t]
    \centering
    \vspace*{-3.0ex}
    \includegraphics[trim=0 0.38cm 0 0, clip, width=0.8\textwidth]{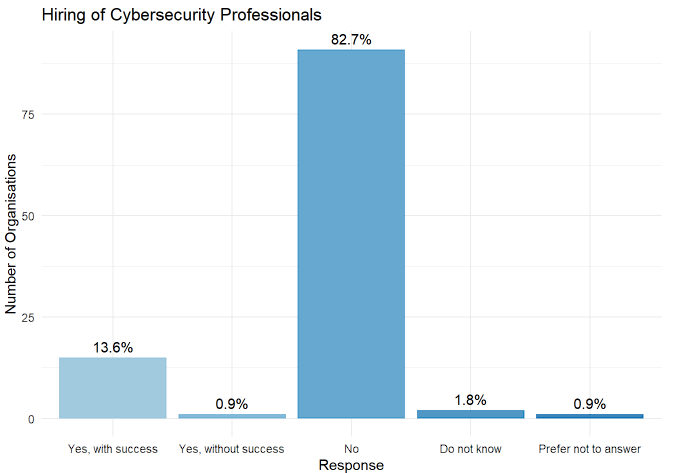}
    \vspace*{-2.0ex}
    \caption{Hiring (Q12.1 Has your organisation hired or attempted to hire
professionals with cybersecurity skills in the past year?)}
    \label{fig:Q12.1 Hiring}
\end{figure}

\begin{figure}[!t]
    \centering
    \vspace*{-2.5ex}
    \includegraphics[width=0.85\textwidth]{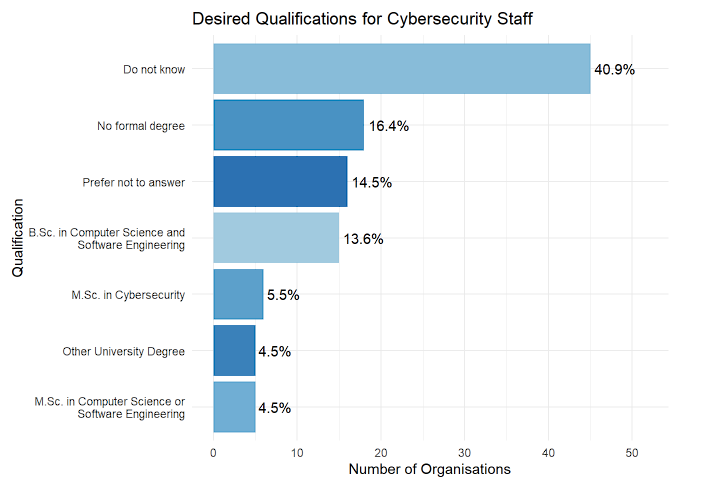}
    \vspace*{-3.0ex}
    \caption{Hiring (Q13 If you hire, what qualification would your staff member responsible for cybersecurity need?)}
    \label{fig:Q13 qualifications}
\end{figure}

The answers to question Q12.1 \quot{Has your organisation hired or attempted to hire professionals with cybersecurity skills in the past year?} are depicted in Figure~\ref{fig:Q12.1 Hiring}. 83\% of the survey respondents expressed that their organisations have not hired or attempted to hire professionals with cybersecurity skills in the ``past year'' (2024/2025), while 14\% hired with success, and only less than 1\% attempted to hire without success. 

When asked (if applicable) the open-ended question, \quot{What challenges has your organisation
faced when recruiting talent for cybersecurity positions?} (Q12.2), respondents most frequently pointed to difficulties in finding staff with sufficient knowledge and experience. Other commonly mentioned challenges included high costs and finding candidates who fit into the organisation's workplace culture. 

Question Q13 asked, \quot{If you hire, what qualification would your staff member responsible for cybersecurity need?}. The largest share (41\%) indicated that they did not know what qualification would be required. 15\% preferred not to answer. 
16\% expressed that no formal degree is needed, while 28\% were in favour of some university degree to qualify for cybersecurity positions in their organisations. For a further breakdown, see Figure~\ref{fig:Q13 qualifications}.

\section{Discussion}\label{sec:discussion}
This section analyses the survey answers and puts them into context with other work. 
The survey data, particularly the results from the open-ended answers (sections~\ref{sec:Cybersecurity challenges and barriers}, \ref{sec:Training employees}, and~\ref{sec:Hiring cybersecurity experts}), show the presence of human behaviour-related factors contributing to cybersecurity difficulties across most of the identified cybersecurity challenges and barriers in Icelandic \acp{SME}. 

Notably, when comparing these results with the related work in Section~\ref{sec:Related Work}, similarities include: a general lack of staff expertise, a false sense of security, managerial overconfidence, as well as ineffective training policies, time and financial resources constraints, and organisational culture.

\subsection{Roles and expertise}\label{sec:Roles and expertise}

Reflecting on the study participants' organisational background, it is important to note that the invitation to participate in the survey was sent to the person in charge of the organisation's cybersecurity. While 66\% of the survey participants reported having IT-specific roles at their organisation (second paragraph of Section~\ref{sec:results}), when asked who is responsible for cybersecurity in their organisations (Q4), 46\% selected the roles Chief Executive Officer (CEO), School master, Director. This suggests that cybersecurity in a large part of respondents' organisations is not the responsibility of professionals who can devote their entire time to cybersecurity. It also indicates that many respondents have multiple roles within their organisations (e.g.\ for a very small \ac{SME}, the CEO is simultaneously in charge of IT) which might negatively affect their cybersecurity qualification. 
Other studies show that IT professionals' cybersecurity awareness levels in \acp{SME} are low because of conflicting priorities and due to lack of appropriate education~\cite{chidukwani2022survey}.

Furthermore, the responses related to hiring (Section~\ref{sec:Hiring cybersecurity experts}) suggest that
there is not a lot of in-house cybersecurity expertise. On the other hand, \acp{SME} in general often hire external companies to take care of their systems~\cite{heidt2019investigating}; hence, recruitment of in-house talent for cybersecurity positions is often not needed. It is important to acknowledge the presence of employees with a foreign background in Icelandic SMEs, especially due to the present need to attach foreign labour in the country arising from the ageing of the Icelandic labour force. Some of the existing literature shows that expatriates easily adjust to Icelandic culture due to quick establishment of trust, and almost non-existent language barriers as the majority of locals speak English~\cite{gudhmunsdottirexpatriate}. The survey data indicate, however, that language factors can still introduce additional layers of challenges for employees with a foreign background when identifying scam emails received in the Icelandic language (Section~\ref{sec:Cybersecurity challenges and barriers}).

The above findings show that \acp{SME} in Iceland are not different from \acp{SME} in other countries with respect to cybersecurity roles and hiring. However, one aspect stands out when compared to other countries, even in comparison to other Nordic countries that are often used as a benchmark for Iceland:
the Icelandic \acp{SME} show a low preference for university degrees as a qualification for hired cybersecurity staff (Section~\ref{sec:Hiring cybersecurity experts}). Moreover, it has to be noted that the largest share of responses was \quot{Do not know}. 
In contrast, international research indicates that employers globally frequently rely on formal university degrees as a primary filtering mechanism and baseline requirement for cybersecurity positions~\cite{ramezan2023examining}.
However, these results are confirmed by other non-cybersecurity labour market research: the Icelandic labour market is characterised by high flexibility and a pragmatic `just do it' work culture~\cite{oladottir2017nordic}. In addition, there is a persistent skills mismatch~\cite{oecd2019fostering}, meaning employers frequently hire based on demonstrated skills and experience rather than relying on formal university degrees.

In fact, this `skills-first' approach utilised by Icelandic \acp{SME} is exactly what the World Economic Forum advocates~\cite{wef2024strategic}, noting that organisations worldwide (specifically including \acp{SME}) struggle to fill cybersecurity positions precisely because they over-rely on traditional educational degrees.

\subsection{False sense of security}

For the surveyed \acp{SME}, there is a clear discrepancy between perceived protection and the reality of their incidents, and security audits (Section~\ref{sec:Perceived protection, real incidents and protection measures}): the organisations feel far better protected than the actual incidents and the lack of protection measures would indicate. This suggests that many respondents may be operating under a false sense of security regarding their cybersecurity posture. 

This aligns with existing literature indicating that \ac{SME} decision-makers frequently exhibit optimism bias and managerial overconfidence, believing their organisations are adequately protected despite lacking empirical evidence~\cite{alahmari2021investigating}. Furthermore, organisations often confuse the implementation of basic IT controls with actual security but without aggressively measuring the effectiveness of these defences (such as through regular penetration testing), \acp{SME} remain unaware of their true vulnerability until a successful cyberattack occurs~\cite{enisa2021cybersecurity,vuggumudi2022false}.  

However, respondents were notably more realistic regarding the threat posed by `shadow IT' (Table~\ref{tab:perceived_protection_threats}). As recent academic literature emphasises, shadow IT creates critical blind spots for organisations, representing a significant insider threat that increases susceptibility to data breaches, compliance violations, and cyberattacks~\cite{haag2024dealing,vanacken2024who}.

To address this false sense of security, \acp{SME} must bridge the gap between perceived and actual protection by elevating the organisation's overall cybersecurity expertise -- for example, through targeted employee training programmes, the strategic hiring of dedicated staff, or, particularly for resource-constrained \acp{SME}, by external cybersecurity service providers.

\subsection{Ineffective training policies}

Employee training was among the most frequently listed challenges (Section~\ref{sec:Cybersecurity challenges and barriers}). However, when respondents were asked whether their organisations offered cybersecurity awareness training (Section~\ref{sec:Training employees}), more than two-thirds answered in the negative.  The responses to the challenges and barriers (Section~\ref{sec:Cybersecurity challenges and barriers}) suggest a reason why training is not offered:
namely, costs associated with training and expressed uncertainty about whether training is effective. 
That data also made clear that training is a challenge due to a lack of interest of employees, which is partly due to inconveniences associated with training, a lack of time, or a lack of a sense of responsibility towards the organisation's cybersecurity.

A number of studies show the need for organisations’ risk management plans to include cybersecurity awareness training to keep organisations secure~\cite{parkar2024cybersecurity,chidukwani2022survey}.  
On the other hand, while security awareness programmes frequently rely on simulated phishing campaigns, recent empirical research indicates that such approaches are often ineffective at producing long-term behavioural change~\cite{11023357} and can even be counterproductive~\cite{lain2022phishing}. Rather than fostering a culture of security, poorly implemented phishing simulations frequently annoy staff, contribute to `cybersecurity fatigue' and organisational mistrust~\cite{brunken2023hidden}. Furthermore, the  nature of these campaigns has been shown to induce unnecessary workplace stress and lower employees' self-efficacy regarding threat detection~\cite{schops2024case}. Therefore, organisations should aim to target training toward their specific needs as opposed to generic solutions which increase stress  \cite{corradini2020building}.

\subsection{Time, staff and financial resource constraints}

In contrast to large organisations, \acp{SME} often experience difficulties in managing workplace stress because of their limited resources~\cite{molek2026current,heidt2019investigating}. The survey data on challenges and barriers (Section~\ref{sec:Cybersecurity challenges and barriers}) show three intersecting resource challenges of \acp{SME} in Iceland that are heavily affected by human factors.

The first resource challenge is the shortage of time; a significant share of the survey respondents mentioned a lack of time to, for example, ensure that employees attend trainings to obtain cybersecurity knowledge or to ensure that the systems are up-to-date. Considering that the biggest challenge (Section~\ref{sec:Cybersecurity challenges and barriers}) is social engineering, such as phishing emails, ensuring that employees have enough cybersecurity awareness knowledge is a crucial task. 
According to Chowdhury et al.,\ time pressure is among the main factors impacting human behaviour and decision-making and it plays a crucial role behind non-secure behaviour~\cite{chowdhury2020time}. 
However, employees' lack of a sense of responsibility towards their organisation's cybersecurity as well as time pressure affecting managerial practices, such as unwillingness to spend money  on up-to-date systems or to hire more staff (Section~\ref{sec:Cybersecurity challenges and barriers}) can be seen as organisational culture challenges.

The second resources-related challenge of Icelandic \acp{SME} is shortage of staff with expertise in cybersecurity (sections~\ref{sec:Cybersecurity challenges and barriers} and~\ref{sec:Hiring cybersecurity experts}). Modern organisations characterised by hyper-connectivity face challenges in supporting their cyber programmes due to a shortage of qualified cybersecurity professionals~\cite{singh2023stress}. The hiring situation (Section~\ref{sec:Hiring cybersecurity experts}) suggests that not only is the availability of such professionals in the labour market an issue for Icelandic \acp{SME}, but the lack of funding for hiring additional staff represents an equally significant barrier.

The third resource-related challenge identified based on the survey data is exactly that lack of funding for cybersecurity (Section~\ref{sec:Cybersecurity challenges and barriers}). Hiring limitations imposed by the lack of financial resources in \acp{SME} are a known challenge, as compared to large organisations, \acp{SME} are less likely to hire dedicated general IT staff, without even considering cybersecurity specialists~\cite{chidukwani2022survey}.

According to Sapiński et al.\,, there is a need to shift the perception of cybersecurity from purely technological to one that takes into consideration not only technology but also psychology~\cite{sapinski2025cyber}.  This idea aligns with Khadka's and Ullah's encouragement to integrate psychological resilience and socio-technical approach to bridge the gap between technical and human factors~\cite{khadka2025human}. Psychological factors can impact cybersecurity decision-making affecting resilience and vulnerability, for instance, stress and cognitive fatigue increase the likelihood of errors~\cite{khadka2025human}. Singh et al.\ point out that cybersecurity professionals experience stress related to unpredictability in job tasks due to ever-changing and advancing security threats~\cite{singh2023stress}. 

Although the survey was completed by a broad range of professionals responsible for digital security within their respective SMEs -- rather than exclusively by dedicated cybersecurity personnel (see Section~\ref{sec:Cybersecurity management role}) -- the data clearly highlighted the challenge of staying up-to-date with newly emerging threats.
Thus, while the stress factor has not been explicitly named by the survey respondents, the challenge of unpredictability, the need for \quot{staying on their toes for a new threat},
lack of resources (time, IT staff, and funding) were listed among cybersecurity challenges -- a list of challenges that have the potential to cause stress.

In summary, staff responsible for cybersecurity in Icelandic \acp{SME} experience some level of stress due to unpredictability and ever-developing security threats. This stress is often heightened by a lack of time and/or funding, outdated systems, but most significantly by a shortage of IT staff -- essential conditions required to address newly emerging cybersecurity threats effectively. 

One approach for SMEs who lack the resources may be external government support~\cite{rawindaran2023perspective}. The European Cybersecurity Competence Centre's network of National Coordination Centres across Europe is therefore a step in the correct direction.

\subsection{Organisational culture}
It is evident from the survey data that some \acp{SME} in Iceland face persisting organisational culture challenges that are resulting not only from employees' reluctance to follow successful cybersecurity practices (Section \ref{sec:Cybersecurity challenges and barriers}) but also as a consequence of managerial practices. For instance, a significant share of written responses included the issue of being understaffed without elaborating too much on why this issue exists; others have noted that it is due to high cost; only a small share of the written answers included more detailed explanations. The fact that the most common word in the open answers for challenges and barriers is \quot{employees} indicates that from the managerial perspective employee behaviour is at fault. 

However, managerial responsibilities were identified as key barriers to effective cybersecurity such as low rates of mandatory cybersecurity awareness training (Section \ref{sec:Training employees}, question Q9.1), low rates of hiring (Section~\ref{sec:Hiring cybersecurity experts}) and a lack of understanding of which qualifications matter. This suggests that changes in managerial culture are key to lowering these barriers. Although it should be reiterated that cybersecurity degrees are not always advised (see the discussion in Section~\ref{sec:Roles and expertise}) and that the efficacy of some training (e.g.,\ phishing) has recently been cast into doubt~\cite{11023357,lain2022phishing,brunken2023hidden}, requiring more targeted approaches. Regardless, these conflicts between employee and employer responsibilities for cybersecurity within the organisations suggests room for improvement in cybersecurity culture, for example, constructive communication (such as a dedicated cybersecurity communication champion~\cite{alshaikh2020developing}) could support defining shared and collaborative responsibilities~\cite{corradini2020building}.

\section{Summary and conclusions}\label{sec:summary}
This study of 130 Icelandic \acp{SME} investigates the influence of human factors on cybersecurity of these organisations from a management perspective. The findings show that in Icelandic \acp{SME}, human factors were indeed influential challenges and barriers to cybersecurity. 

An analysis of these results identified five themes, many of which are aligned with studies in other countries, see related work in Section~\ref{sec:Related Work}: (i) Inadequate capacity of cybersecurity roles and expertise. However, one reason could be related to Iceland's perspective on hiring for skills over formal qualifications, an approach that is also in line with World Economic Forum guidelines. (ii) Organisations having a false sense of security, in that they felt they were adequately protected yet lacked the expertise to be accurately aware of threats to their organisation. Such issues can only be solved by increasing cybersecurity expertise either internally or externally. (iii) The use of ineffective training policies. In line with related literature, cybersecurity training was inconsistently applied, which is concerning given the lack of general cybersecurity expertise. However, we recommend that education policies should be applied in a targeted fashion to prevent employee fatigue. \mbox{(iv) Resource} constraints including time, finances, and staff. These were common for \acp{SME} with limited resources and can  be mitigated through external government support. (v) A general lack of cybersecurity culture creating ambiguity around cybersecurity roles and responsibilities. To mitigate this, we recommend the implementation of a dedicated cybersecurity champion to promote constructive communication within the organisation.  In future work, we aim to extend this study through a deeper multivariate analysis of this data to understand the relationship between the studied factors and incident occurrence. We will also conduct new studies focusing upon the employee (rather than managerial) perspective.

\subsubsection*{Acknowledgments.} 
{\small
This project has received co-funding from the European
Union’s Digital Europe Programme under grant agreements no.\ 101127453 (NCC-IS) and 101226821 (EYVOR NCC-IS), managed by the European Cybersecurity Competence Centre (ECCC). Views and opinions expressed are however those of the authors only and do not necessarily reflect those of the European Union or the ECCC. Neither the European Union nor the granting authority can be held responsible for them.}

\bibliography{bibliography}

\appendix
\section{Statistical Significance Based on Chi-Square Analysis of Survey Data}\label{appendix:chi_square}

This appendix provides details on the statistical significance of the survey data by applying the Chi-Square Goodness-of-Fit test~\cite{agresti2018introduction} to the relevant survey questions covered in this paper. 

Table~\ref{table:chi_square_results} presents the analytical parameters for each non-demographic question with mutually exclusive answer categories. For each item, it provides the number of valid responses ($N$), the degrees of freedom ($df$, derived from the number of answer categories), the calculated $\chi^2$ value, and the applicable critical value at a significance level of $\alpha=0.05$.

As an example, consider the first sub-question of Q6.1 (see Table~\ref{tab:q61 perceived protection} in Section~\ref{sec:Perceived protection, real incidents and protection measures}) that has four discrete answer categories where 19 \quot{Do not know} and \quot{Prefer not to answer} responses are excluded from the total sample of 118 respondents. With the remaining $N=99$ valid responses distributed across observed frequencies of 37, 45, 13, and 4, this question yields a $\chi^2$ value of $45.61$.
In this case, the null hypothesis -- which assumes random answers leading to an equal distribution of responses across all four answer categories -- can be rejected because the $\chi^2$ value substantially exceeds the critical value of $7.81$ that is applicable for four answer categories, i.e.\ 3 degrees of freedom at a significance level of $\alpha=0.05$ (i.e.\ a $5\%$ chance that a result is a false positive). 
This indicates a statistically significant meaningful preference among the respondents, rather than a random
distribution of opinions.

As can be seen in Table~\ref{table:chi_square_results}, $\chi^2$ is 
typically above the critical value. The only exception is question Q14.3 \quot{If yes, what was the attack vector used?} which had rather few responses (the total number of responses was only 25) because it was only presented to those survey participants who had before answered that their organisation experienced a 
cybersecurity incident in the past year (Q14.1).

\begin{table}[!h]
    \centering
    \footnotesize 
    \renewcommand{\arraystretch}{1.2} 
    \setlength{\tabcolsep}{3pt} 
    \caption{Chi-Square Goodness-of-Fit Analysis of Survey Data ($\alpha=0.05$)}
    \label{table:chi_square_results}
    \begin{tabular}{>{\raggedright\arraybackslash}m{1.5cm} *{4}{>{\centering\arraybackslash}m{1.2cm}}}
        \toprule
        \textbf{Survey Question} & 
        \textbf{$N$} & 
        \textbf{$df$} & 
        \textbf{$\chi^2$} & 
        \textbf{Critical Value} \\
        \midrule
        Q6.1.1 &  99 & 3 & 45.61 & 7.81 \\
        Q6.1.2 & 102 & 3 & 49.14 & 7.81 \\
        Q6.1.3 & 100 & 3 & 48.56 & 7.81 \\
        Q6.1.4 & 97  & 3 & 29.8 & 7.81 \\
        Q6.1.5 & 84  & 3 & 61.24 & 7.81 \\
        Q6.1.6 & 102 & 3 & 58.08 & 7.81 \\
        Q6.1.7 & 109 & 3 & 37.97 & 7.81 \\
        Q6.2.1 & 105 & 3 & 16.18 & 7.81 \\
        Q6.2.2 & 103 & 3 & 15.80 & 7.81 \\
        Q6.2.3 & 100 & 3 & 48.56 & 7.81 \\
        Q6.2.4 &  98 & 3 & 10.00 & 7.81 \\
        Q6.2.5 & 103 & 3 & 25.27 & 7.81 \\
        Q6.2.6 & 106 & 3 & 7.89 & 7.81 \\
        Q8 & 97 & 2 & 6.21 & 5.99 \\
        Q9.1 & 106 & 1 & 19.96 & 3.84 \\
        Q12.1 & 107 & 2 & 131.51 & 5.99 \\
        Q13 & 49 & 4 & 15.80 & 9.49 \\
        Q14.1 & 104 & 1 & 28.04 & 3.84 \\
        Q14.3 & 22 & 4 & 8.00 & 9.49 \\ 
        \bottomrule
    \end{tabular}
\end{table}

\end{document}